\documentclass{article}
\usepackage[utf8]{inputenc}
\usepackage{mathtools,halloweenmath}
\usepackage{amsthm}

\begin{document}

\raggedright{\textbf{\large{Monotonicity Failure In Ranked Choice Voting - Necessary and Sufficient Conditions for 3-Candidate Elections}}} \\ 
\medskip{}

\textbf{Rylie E. Weaver} \\
\bigskip{}
\bigskip{}

\textbf{Abstract} Ranked choice voting is vulnerable to monotonicity failure - a voting failure where a candidate is cost an election due to losing voter preference or granted an election due to gaining voter preference. Despite increasing use of ranked choice voting at the time of writing of this paper, the frequency of monotonicity failure is still a very open question. This paper builds on work done by Joseph Ornstein and Robert Norman \cite{ornstein2014} to develop conditions which can be used to test if it's possible that monotonicity failure has happened in a 3-candidate ranked choice voting election. \\
\bigskip{}

\section{Introduction} 
    \hspace{8pt} Ranked choice voting (abbreviated RCV) is a voting system where voters rank candidates in order of preference, and then the candidates are successively eliminated in rounds until just one candidate remains. In each round, the candidate with the least number of first-place votes is who's eliminated. When a candidate is eliminated, all their votes are transferred to their voters' second choice (if a voter's second choice isn't available, then their third choice, etc...). Ranked choice voting is vulnerable to both types of monotonicity failure: upward (when a candidate loses an election due to gaining voter preference) and downward (when a candidates wins an election due to losing voter preference). Certainly, monotonicity failure is not desirable, and depending on its frequency, monotonicity failure could be a cause for concern for those implementing RCV. This paper makes efforts to contribute to our understanding of the frequency of monotonicity failure. \\
    \hspace{8pt} Unfortunately, the election results of a certain election is not enough to determine whether a monotonicity failure has happened. In order to determine whether monotonicity failure has happened or not, one would need the election results at two different times, so that the change, or lack of change, of voter preference and winner could be observed. Further research could approach the question with both polling and election results, but the approach of this paper is to answer a slightly different question: \textbf{Under which elections is it \emph{possible} that monotonicity failure has happened?} By understanding how often it's \emph{possible} that monotonicity failure has happened, we have an upper bound for how often monotonicity failure \emph{does} happen. \\
    \hspace{8pt} For 3-candidate ranked choice voting elections, half of the question has been answered already. Necessary and sufficient conditions have been found for when it's possible to create an \emph{upward} monotonicity failure (reference 1). As this paper will explain later, this is equivalent to when it's been possible that \emph{downward} monotonicity failure has happened. This paper builds on the previous research by finding a necessary and sufficient condition for it to have been possible that upward monotonicity failure has happened. For the rest of this paper, unless specified otherwise, it's assumed that we're only talking about ranked choice voting elections with three candidates.\\

\section{Examples and Background}
    \hspace{8pt} In an RCV election, the candidates can be called $A$, $B$, and $C$, where candidate $A$ is the RCV winner and candidate $C$ is in last place in the first round WLOG. \\
    \hspace{8pt}The results of a ranked choice voting election can be called an election profile, denoted $P$, where $P = (a_1, a_2, b_1, b_2, c_1, c_2)$ from table 1. Column $a_2$, for example, represents the number of of voters who have candidate $A$ as their first preference, candidate $C$ as their second preference, and candidate $B$ as their third preference. Note that this assumes that the each ballot in the RCV election is filled out completely, which is an assumption of this paper. Looking at RCV elections with incomplete ballots could be the subject of future research.\\
    \medskip{}
    
    \hspace{8pt} Define: \\
    \hspace{24pt} (1) $A = a_1 + a_2$, $B = b_1 + b_2$, and $C = c_1 + c_2$ \\
    \hspace{24pt} (2) $X(Y)$ as the number of first place votes $X$ has after the elimination \\
    \hspace{40 pt} of $Y$ for any candidates $X$ and $Y$. For example, $C(B) = C + b_2$ \\
    \hspace{24pt} (3) $V$ as the total number of voters, so $V = A + B + C$ \\
    \medskip{}
    
    \hspace{8pt} We can summarize an RCV election with profile $P$ using arrows and orderings of candidates. For example, if an election profile $P$ has $A$ in first place, $B$ in second place, and $C$ in third place in round one, then $A(C)$ in first place and $B(C)$ in second place in round two, then $A(BC)$ is the final candidate remaining in round three, we could say that $P$ `plays out' as follows: $P: ABC \rightarrow A(C)B(C) \rightarrow A(BC)$. Notice that the right arrows represent a round of the election passing, and the ordering of the candidates from left to right represents the placement from first to last.
    \bigskip{}
    
    \begin{tabular}{|l|l|l|l|l|l|}
        \hline
        $a_1$ & $a_2$ & $b_1$ & $b_2$ & $c_1$ & $c_2$\\
        \hline
        A & A & B & B & C & C \\
        \hline
        B & C & A & C & A & B \\
        \hline
        C & B & C & A & B & A \\
        \hline
    \end{tabular} \hspace{8pt} Table 1 \\
    \bigskip{}

    \hspace{8pt} By changing some voters' preferences, a new election profile can be created, called $P'$, where $P' = (a_1', a_2', b_1', b_2', c_1', c_2')$. \\
    \medskip{}
    
    \hspace{8pt} Define: \\
    \hspace{24pt} (1) $\Delta x_i = x_i - x_i'$ for any candidate and column so that we get \\
    \hspace{42pt} $P' = (a_1 - \Delta a_1, a_2 - \Delta a_2, b_1 - \Delta b_1, b_2 - \Delta b_2, c_1 - \Delta c_1, c_2 - \Delta c_2)$ \\
    \hspace{24pt} (2) $X' = X - \Delta X$ \\
    \hspace{24pt} (3) $X(Y)'$ as the number of first place votes $X'$ has after the elimination \\
    \hspace{24pt} (4) $P \rightarrow P'$ as the change in voter preference needed to change profile $P$ \\
    \hspace{42pt} to profile $P'$. $P \rightarrow P'$ can be though of as the set\\
    \hspace{42pt} $(\Delta a_1, \Delta a_2, \Delta b_1, \Delta b_2, \Delta c_1, \Delta c_2)$ \\
    \medskip{}
    
    \hspace{42pt}of $Y'$ for any candidates $X'$, $Y'$ \\
    \hspace{8pt} Note that an RCV election with profile $P'$ can also be succinctly expressed with arrows and orderings of candidates, such as $P': A'B'C' \rightarrow A(C)'B(C)' \rightarrow A(BC)'$. Now, we take a look at a couple examples of monotonicity failure. \\
    \bigskip{}
    
    \subsection{Upward Monotonicity Failure Example}
    
        \hspace{8pt} Define an election profile $P = (18, 20, 7, 25, 20, 10)$ as seen in table 2. In the first round of $P$, $A = 38$, $B = 32$, and $C = 30$, so $C$ is eliminated. The votes for $C$ are transferred, and in round 2, $A(C) = 58$ and $B(C) = 42$, so $B$ is eliminated. Then, $A$ is the last candidate remaining, so $A$ is the RCV winner. We can say that $P$ plays out as $P: ABC \rightarrow A(C)B(C) \rightarrow A(BC)$.\\
        \bigskip{}
        
        \begin{tabular}{|l|l|l|l|l|l|}
            \hline
            $18$ & $20$ & $7$ & $25$ & $20$ & $10$\\
            \hline
            A & A & B & B & C & C \\
            \hline
            B & C & A & C & A & B \\
            \hline
            C & B & C & A & B & A \\
            \hline
        \end{tabular} \hspace{8pt} Table 2 \\
        \bigskip{}
        
        \hspace{8pt} Define a new election profile $P'$ that's exactly the same as $P$, except that the $7$ voters in $b_1$ now prefer candidate $A$ over candidate $B$ so that they're now part of $a_1'$. This means that $\Delta b_1 = 7$, $\Delta a_1 = -7$, and all other deltas are zero, so the resulting $P' = (25, 20, 0, 25, 20, 10)$ as shown in table 3. Note that $P \rightarrow P'$ is only an increase in voter preference for $A$. \\
            \bigskip{}
        
        \begin{tabular}{|l|l|l|l|l|l|}
            \hline
            $25$ & $20$ & $0$ & $25$ & $20$ & $10$\\
            \hline
            A & A & B & B & C & C \\
            \hline
            B & C & A & C & A & B \\
            \hline
            C & B & C & A & B & A \\
            \hline
        \end{tabular} \hspace{8pt} Table 3 \\
        \bigskip{}
        
        \hspace{8pt} In the first round of $P'$, $A = 45$, $B = 25$, and $C = 30$, so $B$ is eliminated. The votes for $B$ are transferred, and in round 2, $A(B) = 45$ and $C(B) = 55$, so $A$ is eliminated. Then, $C$ is the last candidate remaining, so $C$ is the winner. We can say that $P'$ plays out as $P': A'C'B' \rightarrow C(B)'A(B)' \rightarrow C(AB)'$. \\ 
        \hspace{8pt} In election profile $P$, $A$ is the RCV winner. In the election profile $P'$, $A$ isn't the RCV winner, and $P \rightarrow P'$ is only an increase in voter preference for $A$. Therefore, an upward monotonicity failure has occurred.

    \subsection{Downward Monotonicity Failure Example}
    
        \hspace{8pt} Define an election profile $P = (25, 5, 25, 20, 25, 0)$ as seen in table 4. In the first round of $P$, $A = 30$, $B = 45$, and $C = 25$, so $C$ is eliminated. The votes for $C$ are transferred, and in round 2, $A(C) = 55$ and $B(C) = 45$, so $B$ is eliminated. Then, $A$ is the last candidate remaining, so $A$ is the winner. We can express the election with profile $P$ as $P: BAC \rightarrow A(C)B(C) \rightarrow A(BC)$. \\
        \bigskip{}
        
        \begin{tabular}{|l|l|l|l|l|l|}
            \hline
            $25$ & $5$ & $25$ & $20$ & $25$ & $0$\\
            \hline
            A & A & B & B & C & C \\
            \hline
            B & C & A & C & A & B \\
            \hline
            C & B & C & A & B & A \\
            \hline
        \end{tabular} \hspace{8pt} Table 4 \\
        \bigskip{}
        
        \hspace{8pt} Define a new election profile $P'$ that's exactly the same as $P$, except that the $7$ voters in $b_2$ now prefer candidate $C$ over candidate $B$ so that they're now part of $c_2'$. This means that $\Delta b_2 = 7$, $\Delta c_2 = -7$, and all other deltas are zero, so the resulting $P' = (25, 5, 25, 13, 25, 7)$ as shown in table 5. Note that $P \rightarrow P'$ is only a decrease in voter preference for $B$. \\
            \bigskip{}
        
        \begin{tabular}{|l|l|l|l|l|l|}
            \hline
            $25$ & $5$ & $25$ & $13$ & $25$ & $7$\\
            \hline
            A & A & B & B & C & C \\
            \hline
            B & C & A & C & A & B \\
            \hline
            C & B & C & A & B & A \\
            \hline
        \end{tabular} \hspace{8pt} Table 5 \\
        \bigskip{}
        
        \hspace{8pt} In the first round of $P'$, $A = 30$, $B = 38$, and $C = 32$, so $A$ is eliminated. The votes for $A$ are transferred, and in round 2, $B(A) = 63$ and $C(A) = 37$, so $C$ is eliminated. Then, $B$ is the last candidate remaining, so $B$ is the winner. We can say that $P'$ plays out as $P': B'A'C' \rightarrow B(A)'C(A)' \rightarrow B(AC)'$. \\ 
        \hspace{8pt} In election profile $P$, $B$ isn't the RCV winner. In the election profile $P'$, $B$ is the RCV winner, and $P \rightarrow P'$ is only a decrease in voter preference for $B$. Therefore, a downward monotonicity failure has occurred.

    \subsection{What $P'$ Tells Us}
    
        \hspace{8pt} In the previous subsection, we saw two examples of how a voter profile could be changed so that a monotonicity failure occurs. This tells us that in each of the two voter profile examples, a monotonicity failure can be \emph{created}. However, this paper seeks to answer in what election profiles it's possible that monotonicity failure has happened, not in what election profiles it's possible to create a monotonicity failure. \\
        \medskip{}
        
        \hspace{8pt} In this paper, I use the following definitions: \\
        \hspace{12pt} (1) It's possible that a monotonicity failure has happened for an election \\
        \hspace{30pt} profile $P$ when there exists an election profile $P'$ such that $P' \rightarrow P$ \\
        \hspace{30pt} exhibits a monotonicity failure. \\
        \hspace{12pt} (2) It's possible to create a monotonicity failure for an \\
        \hspace{30pt} election profile $P$ when there exists an election profile $P'$ such that \\
        \hspace{30pt} $P \rightarrow P'$ exhibits a monotonicity failure. \\
        \medskip{}
        
        \hspace{8pt} In this subsection, it will be proven that finding when it's possible to create monotonicity failure and finding when it's possible that monotonicity failure has happened are equivalent questions. If the reader is willing to accept this intuitive relationship, subsection 2.3 can be skipped. \\
        
        \subsubsection{Preliminary Information for Proofs}
        \hspace{8pt} At the start of each proof is an image summary of what the proof will show, where the arrow represents a change in voter opinion and the text above the arrow represents the type of monotonicity failure. The section that follows is the rigorous proof. \\
        \hspace{8pt}As a reminder to the reader, if it's possible to create a monotonicity failure with $P$, we say that there exists a $P'$ such that $P \rightarrow P'$ exhibits a monotonicity failure. If it's possible that a monotonicity failure has happened for $P$, we say that there exists a $P'$ such that $P' \rightarrow P$ exhibits a monotonicity failure. \textbf{This will not be restated in the proofs}. \\
        
        \subsubsection{Possible to Create Upward if and only if It's Possible Downward Has Happened}
        
            \textbf{If Possible To Create Upward Then Possible Downward Has Happened}
            \smallskip{}
            
            \begin{center}
                If $P \xrightarrow{Upward} P'$ then $P' \xrightarrow{Downward} P$
            \end{center}
            
            \hspace{8pt} Given $P$ with RCV winner $X$, let there be a $P'$ such that $P \rightarrow P'$ exhibits an upward monotonicity failure. By definition, $X$ is not the RCV winner for $P'$ and $P \rightarrow P'$ is only an increase in voter preference for $X$. \\
            \hspace{8pt} Take the same voter profile $P'$ from the previous paragraph. As said before, the RCV winner for $P'$ isn't $X$ and the RCV winner for $P$ is $X$. Also, since $P' \rightarrow P$ is the inverse of $P \rightarrow P'$, we know that $P' \rightarrow P$ must only be a decrease in voter preference for $X$. By definition, $P' \rightarrow P$ exhibits a downward monotonicity failure. $\qedsymbol$ \\
            \bigskip{}

            \textbf{If Possible Downward Has Happened Then Possible To Create Upward}
            \smallskip{}
            
            \begin{center}
                If $P' \xrightarrow{Downward} P$ then $P \xrightarrow{Upward} P'$
            \end{center}
            
           \hspace{8pt} Given $P$ with RCV winner $X$, let there be a $P'$ such that $P' \rightarrow P$ exhibits a downward monotonicity failure. By definition, $X$ is not the RCV winner for $P'$ and $P' \rightarrow P$ is only a decrease in voter preference for $X$. \\
            \hspace{8pt} Take the same voter profile $P'$ from the previous paragraph. As said before, the RCV winner for $P'$ is not $X$ and the RCV winner for $P$ is $X$. Also, since $P \rightarrow P'$ is the inverse of $P' \rightarrow P$, we know that $P \rightarrow P'$ must only be an increase in voter preference for $X$. By definition, $P \rightarrow P'$ exhibits an upward monotonicity failure. $\qedsymbol$ \\
            \bigskip{}
            
            In combination, the last two sections of the proof tell us that: \\
            \begin{center}
                \emph{Given an election profile $P$ with RCV winner $X$, it's possible to create an upward monotonicity failure if and only if it's possible that a downward monotonicity failure has happened.}
            \end{center}

        \subsubsection{Can Create Downward if and only if It's Possible Upward Has Happened}
        
            \textbf{If Can Create Downward Then Possible Upward Has Happened}
            \smallskip{}
            
            \begin{center}
                If $P \xrightarrow{Downward} P'$ then $P' \xrightarrow{Upward} P$
            \end{center}
            
            \hspace{8pt} Given $P$ with RCV winner $X$, let there be a $P'$ such that $P \rightarrow P'$ exhibits a downward monotonicity failure. By definition, there is different RCV winner $Y$ for $P'$ and $P \rightarrow P'$ is only a decrease in voter preference for $Y$. \\
            \hspace{8pt} Take the same voter profile $P'$ from the previous paragraph. As said before, the RCV winner for $P'$ is $Y$ and the RCV winner for $P$ is $X$ (not $Y$). Also, since $P' \rightarrow P$ is the inverse of $P \rightarrow P'$, we know that $P' \rightarrow P$ must only be an increase in voter preference for $Y$. By definition, $P' \rightarrow P$ exhibits an upward monotonicity failure. $\qedsymbol$ \\
            \bigskip{}

            \textbf{If Possible Upward Has Happened, Then Can Create Downward}
            \smallskip{}
            
            \begin{center}
                If $P' \xrightarrow{Upward} P$ then $P \xrightarrow{Downward} P'$
            \end{center}
            
            \hspace{8pt} Given $P$ with RCV winner $X$, let there be a $P'$ such that $P' \rightarrow P$ exhibits an upward monotonicity failure. By definition, there is a different RCV winner $Y$ for $P'$ and $P' \rightarrow P$ is only an increase in voter preference for $Y$. \\
            \hspace{8pt} Take the same voter profile $P'$ from the previous paragraph. As said before, the RCV winner for $P'$ is $Y$ and the RCV winner for $P$ is $X$ (not $Y$). Also, since $P \rightarrow P'$ is the inverse of $P' \rightarrow P$, we know that $P \rightarrow P'$ must only be a decrease in voter preference for $Y$. By definition, $P \rightarrow P'$ exhibits a downward monotonicity failure. $\qedsymbol$ \\
            \bigskip{}
            
            In combination, the last two sections of the proof tell us that: \\
            \begin{center}
                \emph{Given an election profile $P$ with RCV winner $X$, it's possible to create a downward monotonicity failure if and only if it's possible that an upward monotonicity failure has happened.}
            \end{center}

        \subsubsection{Conclusion}
        
        Sections 2.3.2 and 2.3.3 tell us the following two statements: \\
        \medskip{}
        
            \hspace{12pt} (1) \emph{Given an election profile $P$ with RCV winner $X$, it's possible to create \\
            \hspace{28pt} an upward monotonicity failure if and only if it's possible that a \\
            \hspace{28pt} downward monotonicity failure has happened.} \\
            \hspace{12pt} (2) \emph{Given an election profile $P$ with RCV winner $X$, it's possible to create \\
            \hspace{28pt} a downward monotonicity failure if and only if it's possible that an \\
            \hspace{28pt} upward monotonicity failure has happened.} \\
            \medskip{}
        
        \hspace{8pt} Since upward and downward monotonicity failure are the only types of monotonicity failure, these statements can be logically combined to state: \\
        \bigskip{}
        
            \begin{center}
                \emph{\textbf{Given an election profile P with RCV winner X, it's possible to create a monotonicity failure if and only if it's possible that a monotonicity failure has happened.}} \\
            \bigskip{}
            \end{center}
            
        \hspace{8pt} We can therefore conclude that (a) `Finding when it's possible to create monotonicity failure' and (b) `Finding when it's possible that monotonicity failure has happened' are equivalent questions, which was what we set out to prove.

    \subsection{Previous Results}
    
        \hspace{8pt} As said in the Abstract, this paper builds on work done by Joseph Ornstein and Robert Norman (Reference 1) to develop conditions which can be used to test if it's possible that monotonicity failure has happened in a 3-candidate RCV election. \\
        \hspace{8pt} Specifically, Ornstein and Norman developed necessary and sufficient conditions for when it's possible to create an upward monotonicity failure, which are listed below. For an election profile $P$ with $V$ voters, assume WLOG that $A$ is the RCV winner and $C$ is in last place in the first round. It's possible to create an upward monotonicity failure if and only if: \\
        \medskip{}
        
        \hspace{12pt} (1) $C + b_2 > \frac{V}{2}$ \\
        \smallskip{}
        \hspace{20pt} and \\
        \smallskip{}
        \hspace{12 pt} (2) $C > \frac{V}{4}$ \\
        \medskip{}
        
        \hspace{8pt} In the following section, this paper develops a necessary and sufficient condition for when it's possible to create a downward monotonicity failure.

\section{Necessary and Sufficient Condition to Create Downward Monotonicity Failure}
        
    \hspace{8pt} In order to show that it's possible to create a downward monotonicity failure for an election profile $P$ with RCV winner $X$, I must find a $P'$ with RCV winner $Y$ different than $X$ where $P \rightarrow P'$ is only a decrease in voter preference for $Y$. \\
    
    \subsection{Narrowing The Problem}
        
        \hspace{8pt} Assume WLOG that for an election profile $P$, $A$ is the RCV winner and $C$ is in last place in the first round. There are two possible options for how $P$ can play out. They are:
        \medskip{}
        
        \hspace{12pt} (1) $P: ABC \rightarrow A(C)B(C) \rightarrow A(BC)$ \\
        \smallskip{}
        \hspace{12pt} (2) $P: BAC \rightarrow A(C)B(C) \rightarrow A(BC)$ \\
        \medskip{}
        
        \hspace{8pt} By enumerating out all the possible ways that $P'$ can play out, we will narrow down the number of ways that monotonicity failure can happen. \\
        
        \subsubsection{For Which Candidates Is Downward Monotonicity Failure Possible?}
        \hspace{8pt} Firstly, notice that it's not possible to create a downward monotonicity failure for $A$ or $C$. It's not possible for $A$ because $A$ is already the winner. It's not possible for $C$ because when the only change is that $C$ loses voter preference, they must stay in last place and always lose in round one. Therefore, we can narrow the problem to creating a downward monotonicity failure for $B$. \\
        
        \subsubsection{Analyzing Option 1}
        \hspace{8pt} With option (1), where the only change in $P \rightarrow P'$ is a decrease in voter preference for $B$, the possible ways that $P'$ can play out are:
        \medskip{}
        
        \hspace{12pt} (1.1) $P': A'B'C' \rightarrow A(C)'B(C)' \rightarrow A(BC)'$ \\
        \smallskip{}
        \hspace{12pt} (1.2) $P': A'C'B' \rightarrow A(B)'C(B)' \rightarrow A(BC)'$ \\
        \smallskip{}
        \hspace{12pt} (1.3) $P': A'C'B' \rightarrow C(B)'A(B)' \rightarrow C(AB)'$ \\
        \smallskip{}
        \hspace{12pt} (1.4) $P': C'A'B' \rightarrow A(B)'C(B)' \rightarrow A(BC)'$ \\
        \smallskip{}
        \hspace{12pt} (1.5) $P': C'A'B' \rightarrow C(B)'A(B)' \rightarrow C(BC)'$ \\
        \medskip{}
        
        \hspace{8pt} The reasoning for the previous statement is lengthy and fairly easy to infer, so it's mostly excluded from this paper. However, one example will be provided: It's not possible for $P'$ to play out as $P': A'B'C' \rightarrow B(C)'A(C)' \rightarrow B(AC)'$ because it's not possible for $B(C)'$ to place higher than $A(C)'$. It's not possible for $B(C)'$ to place higher than $A(C)'$ because $A(C)$ placed higher than $B(C)$ in $P$ and $P \rightarrow P'$ is only a decrease in voter preference for $B$. \\
        \hspace{8pt} None of these sub-options lead to a win for $B$, so none of them lead exhibit a downward monotonicity failure. \\
        
        \subsubsection{Analyzing Option 2}
        \hspace{8pt} With option (2), where the only change in $P \rightarrow P'$ is a decrease in voter preference for $B$, the possible ways that $P'$ can play out are:
        \medskip{}
        
        \hspace{12pt} (2.1) $P': A'B'C' \rightarrow A(C)'B(C)' \rightarrow A(BC)'$ \\
        \smallskip{}
        \hspace{12pt} (2.2) $P': B'A'C' \rightarrow A(C)'B(C)' \rightarrow A(BC)'$ \\
        \smallskip{}
        \hspace{12pt} (2.3) $P': A'C'B' \rightarrow A(B)'C(B)' \rightarrow A(BC)'$ \\
        \smallskip{}
        \hspace{12pt} (2.4) $P': A'C'B' \rightarrow C(B)'A(B)' \rightarrow C(AC)'$ \\
        \smallskip{}
        \hspace{12pt} (2.5) $P': C'A'B' \rightarrow A(B)'C(B)' \rightarrow A(BC)'$ \\
        \smallskip{}
        \hspace{12pt} (2.6) $P': C'A'B' \rightarrow C(B)'A(B)' \rightarrow C(AC)'$ \\
        \smallskip{}
        \hspace{12pt} (2.7) $P': B'C'A' \rightarrow B(A)'C(A)' \rightarrow B(AC)'$ \\
        \smallskip{}
        \hspace{12pt} (2.8) $P': B'C'A' \rightarrow C(A)'B(A)' \rightarrow C(AB)'$ \\
        \smallskip{}
        \hspace{12pt} (2.9) $P': C'B'A' \rightarrow B(A)'C(A)' \rightarrow B(AC)'$ \\
        \smallskip{}
        \hspace{12pt} (2.10) $P': C'B'A' \rightarrow C(A)'B(A)' \rightarrow C(AB)'$ \\
        \medskip{}
        
        \hspace{8pt} The sub-options that lead to a win for $B'$ are sub-options (2.7) and (2.9). These can be grouped together as: $A'$ is eliminated in round one, then $B(A)'$ is ranked over $C(A)'$ in round two, then $B(AC)'$ is the only candidate remaining and $B'$ wins. The short form of this can be summarized as: $P': [B'C']A' \rightarrow B(A)'C(A)' \rightarrow B(AC)'$ where the brackets in $[B'C']A'$ indicate that the placement of $B'$ and $C'$ could be in any order.
        
        \subsubsection{Narrowing The Problem}
        It's possible to create a downward monotonicity failure if and only if there exists a $P'$ such that $P': [B'C']A' \rightarrow B(A)'C(A)' \rightarrow B(AC)'$ (sub-options 2.7 and 2.10) and $P \rightarrow P'$ is only a decrease in voter preference for $B$. We have successfully narrowed the problem.

    \subsection{Necessary and Sufficient Condition}
    
        \hspace{8pt} Now, we focus on finding the necessary and sufficient condition for there to exist $P'$ such that $P': [B'C']A' \rightarrow B(A)'C(A)' \rightarrow B(AC)'$ and $P \rightarrow P'$ is only a decrease in voter preference for $B$. \\
        \hspace{8pt} In this section, ``There exists $P'$ such that $P': [B'C']A' \rightarrow B(A)'C(A)' \rightarrow B(AC)'$ and $P \rightarrow P'$ is only a decrease in voter preference for $B$" may simply be referred to as ``The Statement". Table 1 is repeated below to help the reader. \\
        \bigskip{}
        
        \begin{tabular}{|l|l|l|l|l|l|}
            \hline
            $a_1$ & $a_2$ & $b_1$ & $b_2$ & $c_1$ & $c_2$\\
            \hline
            A & A & B & B & C & C \\
            \hline
            B & C & A & C & A & B \\
            \hline
            C & B & C & A & B & A \\
            \hline
        \end{tabular} \hspace{8pt} Table 1 \\
        
        \subsubsection{Necessary and Sufficient Condition}
        
            The Statement is true if and only if \\
            \smallskip{} 
            $A - C < min(b_2, B - A - 1, B + a_1 - \frac{V}{2} - \delta)$ \\
            \smallskip{} 
            where 
            \[ \delta = \begin{cases} 
              \frac{1}{2} & $V is odd$ \\
              1 & $V is even$
            \end{cases}
            \]
            In this section, this may simply be referred to as ``the condition". \\
            \smallskip{}
            
            The condition combines three inequality states which all must be true, and have colloquial meanings as follows: \\
            \smallskip{}
            \hspace{12pt} (1) $A - C < b_2$ must be large enough to vault $C$ over $A$)\\
            \smallskip{}
            \hspace{12pt} (2) $A - C < B - A - 1$ (the difference between $A$ and $C$ must be less than the difference between $B$ and $A$. Otherwise, $B$ will be eliminated when trying to vault $C$ over $A$)\\
            \smallskip{}
            \hspace{12pt} (3) $A - C < B + a_1 - \frac{V}{2} - \delta$ ($B$ must win after $A's$ elimination)\\
            \medskip{}
        
        \subsubsection{Proof}
        
            \hspace{8pt} $\Rightarrow$ Let the statement be true. We know from the statement that there exists a $P'$ such that: \\
            \medskip{}
            
            \hspace{12pt} (1) The first round ranking of $P'$ is $[B'C']A'$ \\
            \smallskip{}
            \hspace{12pt} (2) The second round ranking of $P'$ is $B(A)'C(A)'$ \\
            \medskip{}
            
            \hspace{8pt} The previous two statements are equivalent to: \\
            \medskip{}
            
            \hspace{12pt} (1) $C' > A'$ \\
            \smallskip{}
            \hspace{12pt} (2) $B' > A'$ \\
            \smallskip{}
            \hspace{12pt} (3) $B(A)' > \frac{V}{2}$ \\
            \medskip{}
            
            \hspace{8pt} Which can be rewritten as: \\
            \medskip{}
            
            \hspace{12pt} (1) $C - \Delta C > A - \Delta A$ \\
            \smallskip{}
            \hspace{12pt} (2) $B - \Delta B > A - \Delta A$ \\
            \smallskip{}
            \hspace{12pt} (3) $B + a_1 - \Delta B - \Delta a_1 > \frac{V}{2}$ \\
            \medskip{}
            
            \hspace{8pt} Then, with some algebra: \\
            \medskip{}
            
            \hspace{12pt} (1) $A - C < \Delta A - \Delta C$ \\
            \smallskip{}
            \hspace{12pt} (2) $-\Delta A + \Delta B < B - A$ \\
            \smallskip{}
            \hspace{12pt} (3) $\Delta B + \Delta a_1 < B + a_1 - \frac{V}{2}$ \\
            \medskip{}
            
            \hspace{8pt} Since $P \rightarrow P'$ is only a decrease in voter preference for $B$, $\Delta A = -\Delta b_1$ and $\Delta C = -\Delta b_2$. For illustration of this concept, take a voter in $b_1$ who decreases their ranking of $B$. Their vote will go to $a_1'$ or $a_2'$. For a voter in $b_2$, their vote will go to $c_2'$ or $c_1'$. Also, note that $\Delta B + \Delta a_1 = \Delta B + \Delta A - \Delta a_2 = \Delta b_1 + \Delta b_2 + \Delta A - \Delta a_2 =$ $= \Delta b_1 + \Delta b_2 - \Delta b_1 - \Delta a_2 = \Delta b_2 - \Delta a_2$. This allows equations (1), (2), and (3) to be rewritten as: \\
            \medskip{}
            
            \hspace{12pt} (1) $A - C < \Delta b_2 - \Delta b_1$ \\
            \smallskip{}
            \hspace{12pt} (2) $2\Delta b_1 + \Delta b_2 < B - A$ \\
            \smallskip{}
            \hspace{12pt} (3) $\Delta b_2 - \Delta a_2 < B + a_1 - \frac{V}{2}$ \\
            \medskip{}
            
            \hspace{8pt} Since the only change in voter preference is a decrease in preference for $B$, $\Delta b_1$ must be non-negative and $\Delta a_2$ must be non-positive, so equations (1), (2), and (3) imply: \\
            \medskip{}
            
            \hspace{12pt} (1) $A - C < \Delta b_2$ \\
            \smallskip{}
            \hspace{12pt} (2) $\Delta b_2 < B - A$ \\
            \smallskip{}
            \hspace{12pt} (3) $\Delta b_2 < B + a_1 - \frac{V}{2}$ \\
            \medskip{}
            
            \hspace{8pt} Since no column in $P'$ can be negative, (because that would imply a negative amount of voters with an opinion) $b_2 - \Delta b_2 \geq 0$, which means that $\Delta b_2 \leq b_2$. Given that, equation (1) implies that $A - C < b_2$, which is part of what we wanted to prove. \\
            \hspace{8pt} Since it's not possible to have a fraction of a voter in RCV, equation (2) can be re-stated as $\Delta b_2 \leq B - A -1$. This can be combined with equation (1) to say that $A - C < B - A - 1$, which is part of what we wanted to prove. \\
            \hspace{8pt} Also, since it's not possible to have a fraction of a voter in RCV, equation (3) can be re-stated as $\Delta b_2 \leq B + a_1 - \frac{V}{2} - \delta$ where $\delta$ is $1$ if $V$ is even and $\frac{1}{2}$ if $V$ is odd. This can be combined with equation (1) to say that $A - C < B + a_1 - \frac{V}{2} - \delta$, which is part of what we wanted to prove. \\
            \smallskip{}
            \hspace{8pt} In conclusion, $A - C < b_2$, $A - C < B - A - 1$, and $A - C < B + a_1 - \frac{V}{2} - \delta$. We can therefore say that: \\
            \bigskip{} 
            $A - C < min(b_2, B - A - 1, B + a_1 - \frac{V}{2} - \delta)$ \\
            \smallskip{} 
            where 
            \[ \delta = \begin{cases} 
              \frac{1}{2} & $V is odd$ \\
              1 & $V is even$
            \end{cases}
            \]
            \raggedleft{\qedsymbol{}}
            \bigskip{}
            
            \raggedright{}
            \hspace{8pt} $\Leftarrow$ Let the condition be true. Choose $P'$ such that the changes in voter preference are $\Delta b_1 = 0$, $\Delta b_2 = A - C + 1$, $\Delta a_1 = 0$, and $\Delta c_1 = 0$. Note that this implies that $\Delta a_2 = 0$ and $\Delta c_2 = -(A - C + 1)$. We know that $\Delta b_2 = A - C + 1$ is allowed because $A - C < b_2$ from the statement. Also, note that $P \rightarrow P'$ is only a decrease in voter preference for $B$. Now, we'll show that: \\
            \smallskip{}
            
            \hspace{12pt} (1) $C' > A'$ \\
            \hspace{12pt} (2) $B' > A'$ \\
            \hspace{12pt} (3) $B(A)' > \frac{V}{2}$\\
            \smallskip{}
            
            \hspace{8pt} Assume that $C' > A'$ is not true. This would mean that $C' \leq A'$, which can be rewritten as $C - \Delta C \leq A - \Delta A$. Then, substitute $-\Delta C = \Delta b_2 = A - C +1$ and $-\Delta A = \Delta b_1 = 0$ to get $C + (A - C + 1) \leq A$. This simplifies to $A + 1 \leq A$, which is a contradiction. Therefore, $C' > A'$ is true. \\
            \hspace{8pt} Assume that $B' > A'$ is not true. This would mean that $B' \leq A'$, which can be rewritten as $B - \Delta B \leq A - \Delta A$. Then, substitute $-\Delta B = -\Delta b_1 - \Delta b_2 = -(A - C +1)$ and $-\Delta A = \Delta b_1 = 0$ to get $B - (A - C + 1) \leq A$. This simplifies to $B - A - 1 \leq A - C$, which is a contradiction with the condition. Therefore, $B' > A'$ is true. \\
            \hspace{8pt} Assume that $B(A)' > \frac{V}{2}$ is not true. This would mean that $B(A)' < \frac{V}{2} + \delta$, which can be rewritten as $B + a_1 - \Delta B - \Delta a_1 < \frac{V}{2} + \delta$. Now, substitute $- \Delta B - \Delta a_1$ as $-(A - C + 1)$ to get $B + a_1 - (A - C + 1) < \frac{V}{2} - \delta$, which can be simplified and rearranged to $B + a_1 - \frac{V}{2} - 1 + \delta < A - C$. Now, notice that $-1 + \delta \geq -\delta$, so $B + a_1 - \frac{V}{2} - 1 + \delta < A - C$ implies that $B + a_1 - \frac{V}{2} - \delta < A - C$, which is a contradiction with the condition. Therefore, $B(A)' > \frac{V}{2}$ is true. \\
        
            \hspace{8pt} The results of the previous three paragraphs tell us that: \\
            \smallskip{}
            \hspace{20pt} (1) $C' > A'$ \\
            \hspace{20pt} (2) $B' > A'$ \\
            \hspace{20pt} (3) $B(A)' > \frac{V}{2}$ \\
            \medskip{}
            \hspace{8pt} Given these three results, $P'$ must play out as $P': [B'C']A' \rightarrow B(A)'C(A)' \rightarrow B(AC)'$. Therefore, there exits a $P'$ such that $P': [B'C']A' \rightarrow B(A)'C(A)' \rightarrow B(AC)'$ and $P \rightarrow P'$ is only a decrease in voter preference for $B$. \qedsymbol{}

            \section{Conclusion}
            \hspace{8pt} In conclusion, this paper has done two things: \\
            
            \smallskip{}
            \hspace{12pt} (1) Proved that it's possible to create a monotonicity failure with a voter \\
            \hspace{28pt} profile $P$ if and only if it's possible that monotonicity failure has \\
            \hspace{28pt} happened. \\
            \hspace{12pt} (2) Proved a necessary and sufficient condition for it to be possible to \\
            \hspace{28pt} create a downward monotonicity failure in a 3-candidate RCV election. \\
            \smallskip{}
            
            \hspace{8pt} In conjuction with the results provided by Ornstein and Norman (Reference 1), we have complete conditions to test if it's possible that monotonicity failure has happened for a 3-candidate RCV election. \\
            \hspace{8pt} Future research could look at finding necessary and sufficient conditions for RCV elections with incomplete ballots and/or more than three candidates. \\

\bibliographystyle{plain}  
\bibliography{references}

\begin{thebibliography}{1}

\bibitem{ornstein2014}
J.T. Ornstein and R.Z. Norman.
\newblock Frequency of monotonicity failure under instant runoff voting:
  estimates based on a spatial model of elections.
\newblock {\em Public Choice}, 161:1--9, 2014.

\end{thebibliography}

\end{document}